\documentclass[superscriptaddress,
prc,
twocolumn,
preprintnumbers,
amsmath,
amssymb,
altaffilletter,
floatfix,
]{revtex4-2}

\usepackage{graphicx}
\usepackage{dcolumn}
\usepackage{bm}
\usepackage{caption}
\usepackage{subcaption}

\newcommand{\vbb}{$0 \nu \beta \beta $}
\newcommand{\vvbb}{$2 \nu \beta \beta $}
\newcommand{\C}{$^{10}$C}
\newcommand{\Xe}{$^{136}$Xe}

\newcommand{\Bi}{$^{214}$Bi}

\begin{document}

\preprint{APS/123-QED}

\title{KamNet: An Integrated Spatiotemporal Deep Neural Network for Rare Event Search in KamLAND–Zen}
\thanks{Code Repository: \url{https://github.com/aobol/KamNet.git}}
\newcommand{\unc}{Department of Physics and Astronomy, University of North Carolina, Chapel Hill, NC 27514, USA;\\
and Triangle Universities Nuclear Laboratory, Durham, North Carolina 27708, USA}
\newcommand{\lnsmit}{Laboratory of Nuclear Science, Massachusetts Institute of Technology, Cambridge, MA 02139, USA}
\newcommand{\bu}{Department of Physics, Boston University, Boston, MA, 02215}
\newcommand{\tohoku}{Research Center for Neutrino Science,Tohoku University, Sendai 980-8578, Japan}
\author{A.~Li}~\email{Corresponding Author. Email: liaobo77@ad.unc.edu}\affiliation{\unc}
\author{Z.~Fu}\affiliation{\lnsmit}
\author{C.~Grant}\affiliation{\bu}
\author{H.~Ozaki}\affiliation{\tohoku}
\author{I.~Shimizu}\affiliation{\tohoku}
\author{H.~Song}\affiliation{\bu}
\author{A.~Takeuchi}\affiliation{\tohoku}
\author{L.A.~Winslow}\affiliation{\lnsmit}

\date{\today}

\begin{abstract}
Rare event searches allow us to search for new physics at energy scales inaccessible with other means by leveraging specialized large-mass detectors. Machine learning provides a new tool to maximize the information provided by these detectors. The information is sparse, which forces these algorithms to start from the lowest level data and exploit all symmetries in the detector to produce results. In this work we present KamNet which harnesses breakthroughs in geometric deep learning and spatiotemporal data analysis to maximize the physics reach of KamLAND–Zen, a kiloton scale spherical liquid scintillator detector searching for \vbb. Using a simplified background model for KamLAND we show that KamNet outperforms a conventional CNN on benchmarking MC simulations with an increasing level of robustness. Using simulated data, we then demonstrate KamNet's ability to increase KamLAND–Zen's sensitivity to \vbb~and \vvbb~Decay to excited states. A key component of this work is the addition of an attention mechanism to elucidate the underlying physics KamNet is using for the background rejection.

\end{abstract}

\maketitle


\section{Introduction}
Rare event searches provide a unique window on processes happening at energy scales beyond those currently accessible with accelerators up to and including the GUT-scale. These experiments do this with highly customized detectors to reduce background and large masses to maximize exposure. By their nature the data coming from these detectors is sparse and algorithms for analyzing this data must maximize the available information. This is a natural application for machine learning but a different optimization from the big data applications that have been the main focus of the field.

Monolithic kiloton-scale liquid scintillator (LS) detectors like KamLAND exemplify this approach and have been the work horse of neutrino physics for many decades \cite{KamPRL,KamLAND_reactor,Borexino_CNO,Borexino_7Be1,Borexino_7Be2,snop_nd,snop_solar,dayabay,juno_summary}. 
In the first phase of KamLAND, 1\,kiloton of LS is contained in a 13-m-diameter balloon and this LS-filled balloon surrounded by mineral oil (acting as a buffer volume) is viewed by 1879 photomultiplier tubes (PMTs). The KamLAND–Zen experiment inherits the infrastructure of the KamLAND detector and deploys 24 tons of Xe-loaded liquid scintillator (XeLS) in a 3.80-m-diameter spherical inner-balloon at the center of the KamLAND detector. 

This inner-balloon currently contains XeLS doped with 745$\pm$3\,kg of Xe to search for neutrinoless double-beta decay (\vbb)~\cite{gando2020first}. Fig.~\ref{schematic_plot} (Center) shows the KamLAND–Zen detector.  

The first observation of \vbb~ would prove that the neutrino is its own antiparticle, also known as a Majorana particle.  This is a key ingredient for Leptogenesis~\cite{leptogenesis}, which describes the observed matter-antimatter asymmetry in our universe. The challenge of searching for \vbb~decay with monolithic LS detectors is that they function primarily as calorimeters and lack the sophisticated tracking and topological information provided by other technologies like Time Projection Chambers, Cherenkov ring imaging detectors, or silicon strip trackers.  Enhancing monolithic LS detectors with the capability to discriminate between different event types based on tracking and topology would be a revolutionary advancement.

In this work, our goal is to provide a template for developing ML that makes use of all detector symmetries and information encoded in the low-level data. A key component of this work is the development of tools to interrogate the algorithm to discover which data is key to its performance. The resulting algorithm is KamNet. It is built upon our experience with a conventional CNN~\cite{nim_paper}. However, the conventional CNN lacks the ability to harness certain symmetries embedded in KamLAND–Zen data which suppresses the performance of neural network. 

This state-of-the-art machine learning algorithm harnesses the symmetry of a spherical detector to discriminate between topological differences in energy deposits.
With the power to discriminate between different event topologies, we then apply KamNet to different studies from monitoring data quality to enabling new analyses and of course reducing backgrounds in the \vbb~analysis. The paper is structured as follows. The first section explains the construction of the algorithm: Section~\ref{sec:data} describes the low-level information encoded in the KamLAND–Zen data, Section~\ref{section:simulation} describes the different simulations used, and Section~\ref{sec:network_design} describes the network design. Section~\ref{sec:vbb_decay} outlines the application to the \vbb~analysis. We highlight here KamNet's ability to self-interpret its decision via an attention mechanism, thus unveiling what underlying physics is driving the discrimination power. Section~\ref{sec:excitedstates} introduces a future analysis that uses KamNet to extract the \vvbb~decay to excited states signal for the first time. Finally, in Section~\ref{subsec:interpret} we highlight KamNet's ability to self-interpret its decision via an attention mechanism, thus unveiling what underlying physics is driving the discrimination power. We highlight Section~\ref{subsec:interpret} which presents the application of an attention mechanism to determine what low-level information is driving the discrimination power of KamNet.

\section{KamLAND–Zen Data}
\label{sec:data}

\begin{figure*}[!ht]
    \centering
    \includegraphics[width=1.0\linewidth,,trim={0 8pc 0 10pc},clip]{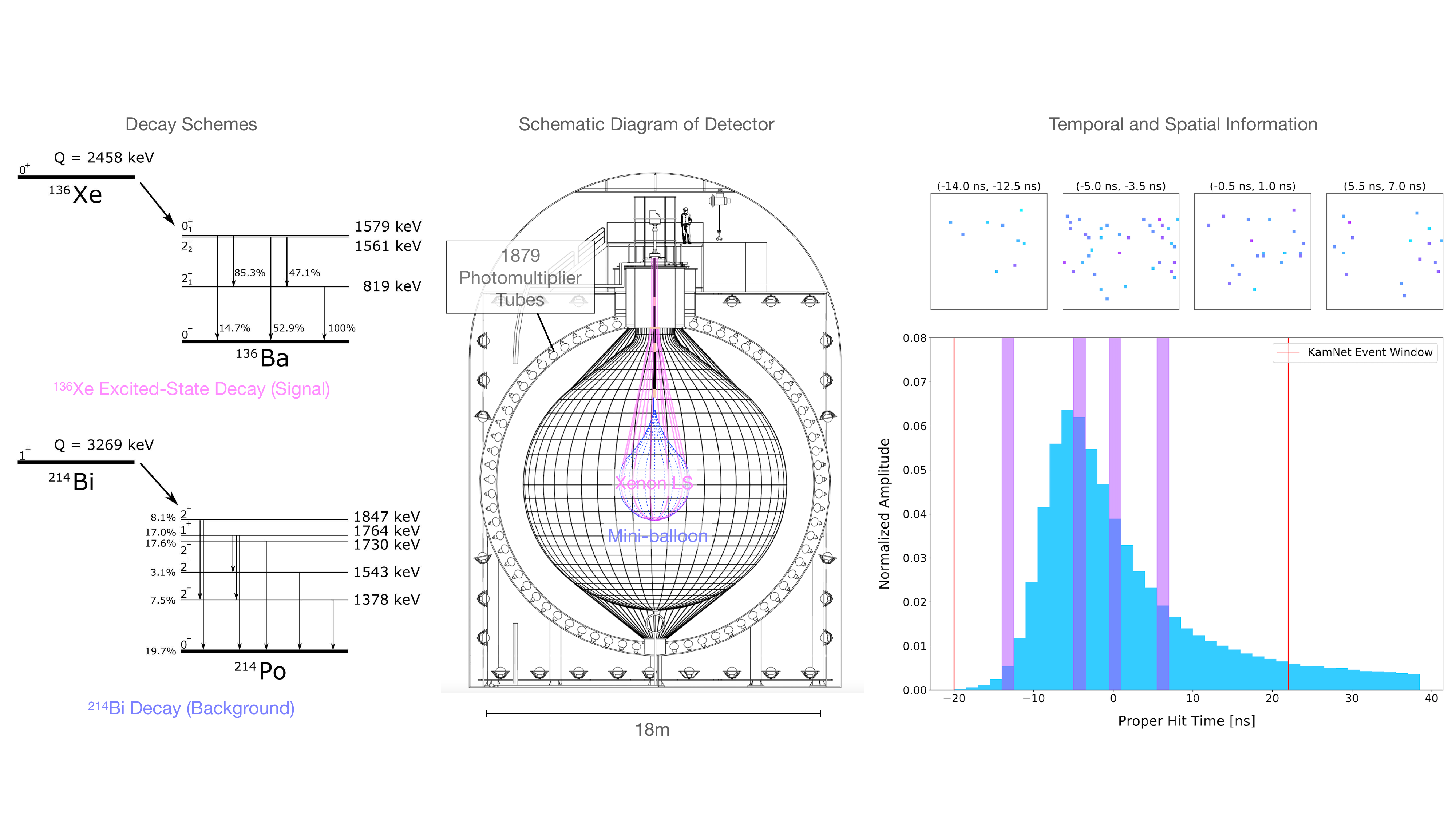}
    \caption{(Left) The decay schemes for $^{136}$Xe and $^{214}$Bi with branching ratios $<1\%$ are omitted for simplicity. (Center) The schematic diagram of the KamLAND–Zen detector. (Right) The distribution of PMT hit times for typical physics events. The spatial distribution of the PMT hit times highlighted in violet are shown above.}
    \label{schematic_plot}
\end{figure*}



The level diagram for double-$\beta$ decay and a common background the single $\beta$ decay of \Bi~are shown in Fig.~\ref{schematic_plot} (Left). These decays can proceed directly to the ground state where the $\beta$ or $\beta$s carry away all of the energy or via an excited state where one or more de-excitation $\gamma$s are emitted. In LS, $\beta$s deposit their energy in a very localized region, resulting in highly isotropic scintillation light. We define this type of event as a single vertex event. In comparison, $\gamma$s can scatter multiple times with a mean free path in the scintillator on the order of $\sim$10\,cm. This multi-site energy deposition results in a slightly less isotropic scintillation light emission and thus we define this type as closely-spaced multi-vertex events.

Each event depositing energy in the XeLS (or LS) produces isotropic scintillation light accompanied by a relatively small amount Cherenkov light.  These photons propagate throughout the detector volume and eventually register hits on a subset of the 1879 PMTs. Among them, 1325 PMTs are 17-inch PMTs with fast timing resolution and contribute 22\% of photo-coverage; the rest are 20-inch PMTs with relatively poorer timing resolution and contribute an additional 12\% photo-coverage. In this way, the raw data for a single event is made up of PMT hit times and a spherical map of the positions of hit PMTs, see Fig.~\ref{schematic_plot} (Right).  The raw data is sparse, but it encodes key information about the underlying physics that can be utilized more efficiently by KamNet compared to traditional cut-based data analyses to disentangle strictly single-vertex events and closely-spaced multi-vertex events in LS detectors.

\section{Simulation Production}
\label{section:simulation}



Three different detector Monte Carlo~(MC) simulations are used to study the performance of KamNet.  Two of the simulations are written using the Reactor Analysis Toolkit (RAT)~\cite{RAT_LOI}.  RAT is a simulation and analysis package that acts as an interface to GEANT4~\cite{geant4one, geant4two}. The first is a simple KamLAND–Zen 400 simulation, referred to as sim-Fast, which is used for very fast benchmarking studies described in Section~\ref{subsec:benchmarking_dataset}.  The second simulation is based on RAT, referred to as sim-RAT, which contains a more detailed model of the KamLAND–Zen 800 detector and is used to accurately characterize KamNet's classification performance for \Xe~excited-state decays in KamLAND–Zen 800. Finally, the performance of KamNet on \vbb~is studied using the official KamLAND–Zen 800 detector MC, which is written purely in GEANT4 and has been carefully tuned to replicate the response of the real detector.  The official simulation for KamLAND–Zen 800, referred to as sim-KLZ800, is quite resource-intensive compared to sim-RAT, which is why it hasn't yet been used for the excited-state study. However, we observe similar classification performance when KamNet is applied to sim-RAT and sim-KLZ800, so we have strong reason to believe the KamNet results for the \Xe~excited-state decays in sim-RAT will carry over when it's applied to the real KamLAND–Zen 800 detector data.  

Both sim-KLZ800 and sim-RAT incorporate the detailed geometry of KamLAND–Zen 800, from the innermost XeLS mini-balloon, out to the 1879 PMTs encased inside the 18-meter-diameter stainless steel sphere as shown in the center of Figure~\ref{schematic_plot}. The optical properties of all the inner detector materials, such as transparency, index of refraction, and reflectivity, are chosen to match those of the KamLAND–Zen detector. The scintillation and Cherenkov photon production is modeled using bench-top measurements of the index of refraction, light yield, absorption, re-emission, and quenching of XeLS and LS as input. After production, all optical photons are allowed to propagate throughout the inner detector volume until they are absorbed. The one difference between the simulations is that sim-KLZ800 has turned off Cherenkov photon production to increase performance without decreasing the MC-data agreement.

PMT photocathodes that absorb optical photons, accompanied by the production of one or more photoelectrons, are referred to as being 'hit'. The time at which a PMT is hit by a photon, $\mathrm{T}_{\mathrm{raw}}$, is measured from the first PMT hit. In KamLAND–Zen, two corrections are applied to the raw hit time of each PMT to give what's called the proper hit time $\tau$. The proper hit time of each PMT is calculated as follows:
\begin{equation}
    \tau = \mathrm{T}_{\mathrm{raw}} - \mathrm{TOF}-\mathrm{T}_{0}
    \label{eqn:PMT_proper_time}
\end{equation}
where TOF is the photon time-of-flight from the event vertex to PMT position and $\mathrm{T}_{0}$ is referred to as the proper start time. The event vertex, used to calculate the TOF, is reconstructed using the standard centroid fitter in RAT. By subtracting $\mathrm{TOF}$ from $\mathrm{T}_{\mathrm{raw}}$, we effectively move the vertex of each event to the center of the detector and correct for intra-event distortion of the scintillation time profile by the vertex position. 
The correction for intra-event distortion of the time profile due to varying energy deposits comes from subtracting $\mathrm{T}_{0}$, which is a fractional charge weighted sum of the differences between $\mathrm{T}_{\mathrm{raw}}$ and TOF over all the PMTs. This is calculated as follows:
\begin{equation}
    \mathrm{T}_{0}=\frac{\sum_{i}(\mathrm{T}_{\mathrm{raw}}^{i} - \mathrm{TOF}^{i})\times \mathrm{Q}_{i}}{\sum_{i}\mathrm{Q}_{i}}
    \label{eqn:PMT_t0}
\end{equation}
\noindent where $Q_i$ is the charge on the $i^\mathrm{th}$ PMT. The proper hit time is calculated inside an interval from -20\,ns to 22\,ns and is binned in increments of 1.5\,ns in order to match the sampling time interval of the KamLAND–Zen readout electronics. Each proper hit time bin contains PMT hits that register photoelectron charge to a $38\times38$ spherical grid segmented in the azimuthal and polar angles, $\theta$ and $\phi$.  The resulting dimensions of the collection of spatiotemporal hit maps is $28\times38\times38$ in $t$, $\theta$, and $\phi$, respectively.  A few slices of a typical spatiotemporal hit map, where the spherical hit maps are projected onto a 2D plane, are shown in the rightmost plot of Figure~\ref{schematic_plot}.


The energies, times, directions, and angular distributions of particles emitted during \vvbb~\Xe~ decay to ground state, \vvbb~\Xe~decay to excited states ($0^+\rightarrow0^+_1$, $0^+\rightarrow2^+_1$, and $0^+\rightarrow2^+_2$), and other radioactive decay backgrounds, are simulated using the standard Decay0 event generator~\cite{decay0}. The simulated events are distributed uniformly within the roughly 3.8-meter-diameter mini-balloon spherical volume at the center of KamLAND–Zen detector.



\section{Network Design}
\label{sec:network_design}

\begin{figure*}[hbt!]
    \begin{subfigure}{0.66\linewidth}
      \centering
      \includegraphics[width=1.0\linewidth,trim={0pc 0pc 0pc 0pc},clip]{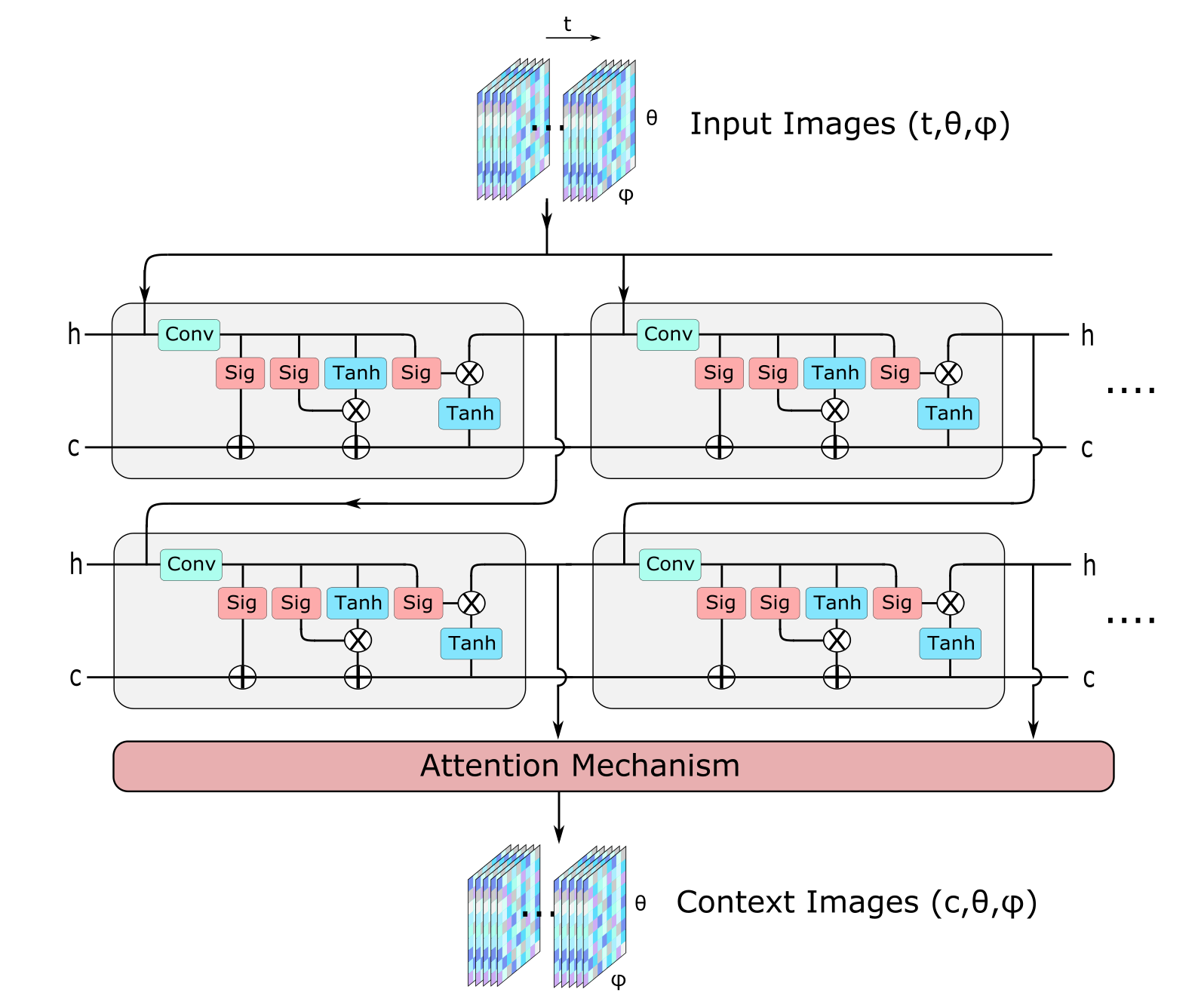}
      \caption{}
      \label{fig:nn_structure_convlstm}
    \end{subfigure}
    \begin{subfigure}{0.33\linewidth}
      \centering
      \includegraphics[width=1.0\linewidth,trim={1pc 0pc 2pc 0pc},clip]{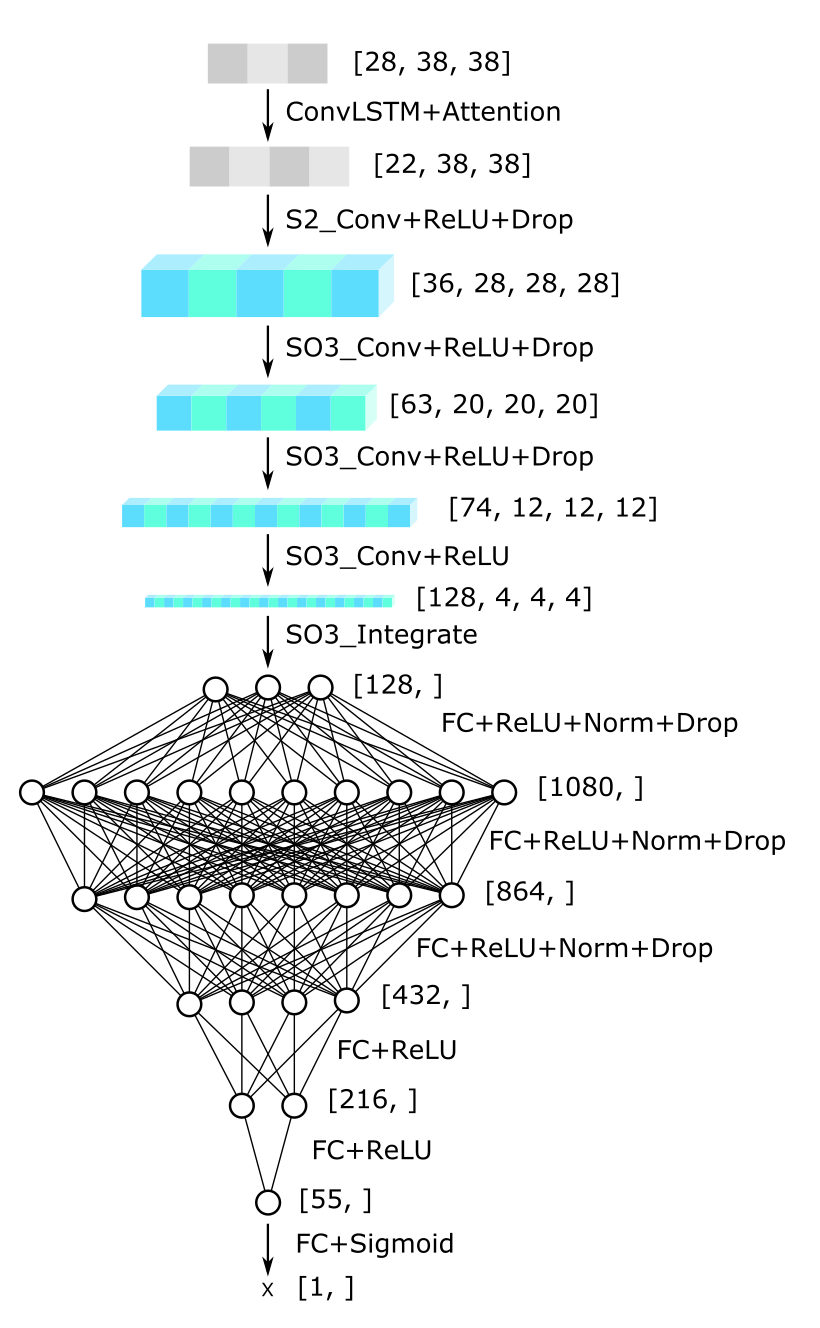}
      \caption{}
      \label{fig:nn_structure_kamnet}
    \end{subfigure}
    \caption{(a) Schematic Diagram of the AttentionConvLSTM layer, including 2 ConvLSTM layers and attention mechanism. (b) Diagram of KamNet.}
    \label{fig:schematic_plot}
\end{figure*}

In our first attempt to apply machine learning to a KamLAND-like detector, we used a conventional CNN to study its ability to reject muon spallation background. We achieved a rejection efficiency of 61\% for \C~while preserving 90\% of \Xe~\cite{nim_paper}.  However, the conventional CNN was originally designed for 2D planar images. Since KamLAND-Zen data is effectively a time series of spherical images, the conventional CNN lacks the ability to harness certain embedded symmetries which suppresses the neural network performance. In this paper, we fully redesigned our deep learning model to recognize the additional symmetries. The new model is referred to as KamNet since it is originally designed for KamLAND–Zen.

KamNet builds on our initial work and a complementary recurrent neural network (RNN) algorithm that was also designed for KamLAND–Zen~\cite{KamRNN}. It is inspired by two recent breakthroughs in geometric deep learning based spherical analysis:  S2CNN~\cite{sphcnn}, and ConvLSTM~\cite{convlstm}.  S2CNN uses a group theory approach to handle spherical data input and ConvLSTM provides a mechanism for understanding time correlations among images.  

\subsection{Rotational Symmetry}
The conventional CNN scans rectangular filters throughout 2D images in a translation-invariant manner, and the translational invariance makes it highly efficient in analyzing planar images. KamLAND–Zen produces spherical PMT hit maps not rectangular images. Mapping the spherical surface of the detector image onto a 2D planar grid will necessarily introduce distortion, which breaks the SO(3) symmetry of spherical signal. In other words, the conventional CNN fails to preserve the rotational invariance of a spherical signal.

The spherical CNN~\cite{sphcnn} is introduced to address this issue. This model uses a group theory approach to incorporate rotational symmetry. In the spherical CNN, the input images undergoes a spherical Fourier transform and the kernels undergo an SO(3) Fourier transform. Both Fourier transforms outputs a 3D representation tensor in Euler angle space. The convolution is performed by multiplication within the Euler angle space to produce a feature map. Each cell in the feature map represents a global convolution between the entire image and filter viewed at a given Euler angle. By scanning the range of all possible rotation angles, the spherical CNN is able to harness the rotational invariance of KamLAND–Zen data. 

\subsection{Temporal Symmetry}
The other key symmetry is temporal symmetry. The input data contains a series of spherical hit maps segmented by time. In the conventional CNN, and spherical CNN, multiple input images are treated as channels. This treatment does not preserve the order of input, and it also doesn't properly handle the short- and long-range temporal correlations.  A Convolutional LSTM (ConvLSTM) model is introduced to resolve this issue~\cite{convlstm}.

Long Short-Term Memory (LSTM) is a classical recurrent neural network model for time series analysis~\cite{lstm}. The LSTM layer contains a hidden state and a cell state to store short-term and long-term information, respectively. The ConvLSTM~\cite{convlstm} combines LSTM with a convolution to analyze the time-ordered images, or spatiotemporal data.  The convolution operation allows the algorithm to understand the spatial correlation within each image, while the LSTM structure allows it to understand short-term and long-term correlations between images at different times.

Generally, the ConvLSTM is sufficient for handling temporal symmetry in a data set, but the inclusion of the ConvLSTM into the spherical CNN model is not so straightforward. The problem originates from the timing profile of scintillation light shown in Figure~\ref{schematic_plot}. Most of the PMT hits are registered within a few spherical hit maps near the scintillation peak, while all other hit maps contains few or zero PMT hits. Therefore, hit maps near the scintillation peak carry much more information than those far from the peak and more weight should be put on the hit maps containing most of the data. In order to accomplish this, an attention mechanism~\cite{attention} is added to ConvLSTM layers and, as far as we know, the method described below is a unique development among the deep learning community. 

When a series of spherical hit maps $\mathrm{I}_{\mathrm{in}}(t,\theta,\phi)$ are fed into the ConvLSTM layers, we obtain two tensors.  The first is the   collection of output images of the intermediate hidden state, denoted $\mathrm{I}_{\mathrm{hidden}}(c, t, \theta, \phi)$, and the second is the final output state of the ConvLSTM, denoted $\mathrm{I}_{\mathrm{output}}(c, \theta, \phi)$.
The time indices are denoted by $t$ and the channels of image are denoted by $c$. Next, the $\mathrm{I}_{\mathrm{output}}(c, \theta, \phi)$ tensor is expanded with a singleton in the time dimension and is used to calculate the attention score $\mathrm{S}(t)$.  The attention score is a weighting factor calculated for each time index as:
\begin{multline}
\mathrm{S}(t) = \mathrm{Softmax}[\,\mathrm{I}_{\mathrm{hidden}}(c,t,\theta,\phi) \circ \mathrm{W}(c,t,\theta,\phi) \\ \circ \mathrm{I}_{\mathrm{output}}(c,\theta,\phi)\,]
\label{eqn:attention_score}
\end{multline}
where $\mathrm{W}(c,t,\theta,\phi)$ is the attention weight matrix learned during training.  The multiplication of the three tensors $\mathrm{I}_{\mathrm{hidden}}(c,t,\theta,\phi)$, $\mathrm{W}(c,t,\theta,\phi)$, and $\mathrm{I}_{\mathrm{output}}(c,\theta,\phi)$ in Equation~\ref{eqn:attention_score} is performed element-wise in a manner equivalent to the Hadamard product of matrices, and the Softmax function is performed along the time dimension. Finally, a context tensor $\mathrm{I}_{\mathrm{context}}(c,\theta,\phi)$ is obtained by:
\begin{equation}
    \mathrm{I}_{\mathrm{context}}(c,\theta,\phi) = \sum_{t}\mathrm{S}(t)\mathrm{I}_{\mathrm{hidden}}(c,t,\theta,\phi)
    \label{eqn:attn_out}
\end{equation}
Thus, the input spatiotemporal hit maps $\mathrm{I}_{\mathrm{in}}(t,\theta,\phi)$ indexed by time are converted to context images $\mathrm{I}_{\mathrm{context}}(c,\theta,\phi)$ indexed by channel. The content in each channel is controlled by the attention score $\mathrm{S}(t)$, which allows the machine to zoom in to important time slices and leave out less important ones. In other words, each channel of the context image is a 2D representation of the spatiotemporal hit map, taking into consideration the temporal symmetry among different hit maps. The model that produces context images is called the AttentionConvLSTM layer, a schematic diagram of this layer is shown in Figure~\ref{fig:nn_structure_convlstm}.

After combining the AttentionConvLSTM layer and spherical CNN, we obtain the full KamNet model shown in Figure~\ref{fig:nn_structure_kamnet}. The input hit maps are first processed by the AttentionConvLSTM layer to produce 32 context images. The context images are then analyzed by the spherical CNN to produce feature vectors. Eventually, the feature vectors are classified by a fully connected network to produce a KamNet score for each event. The input event is more signal-like if KamNet assigns a high attention score, and vice versa. 

\subsection{Benchmarking Dataset}\label{subsec:benchmarking_dataset}
\begin{figure}[hbt!]
    \begin{subfigure}{1.0\linewidth}
      \centering
      \includegraphics[width=1.0\linewidth,trim={12pc 2pc 8pc 0pc},clip]{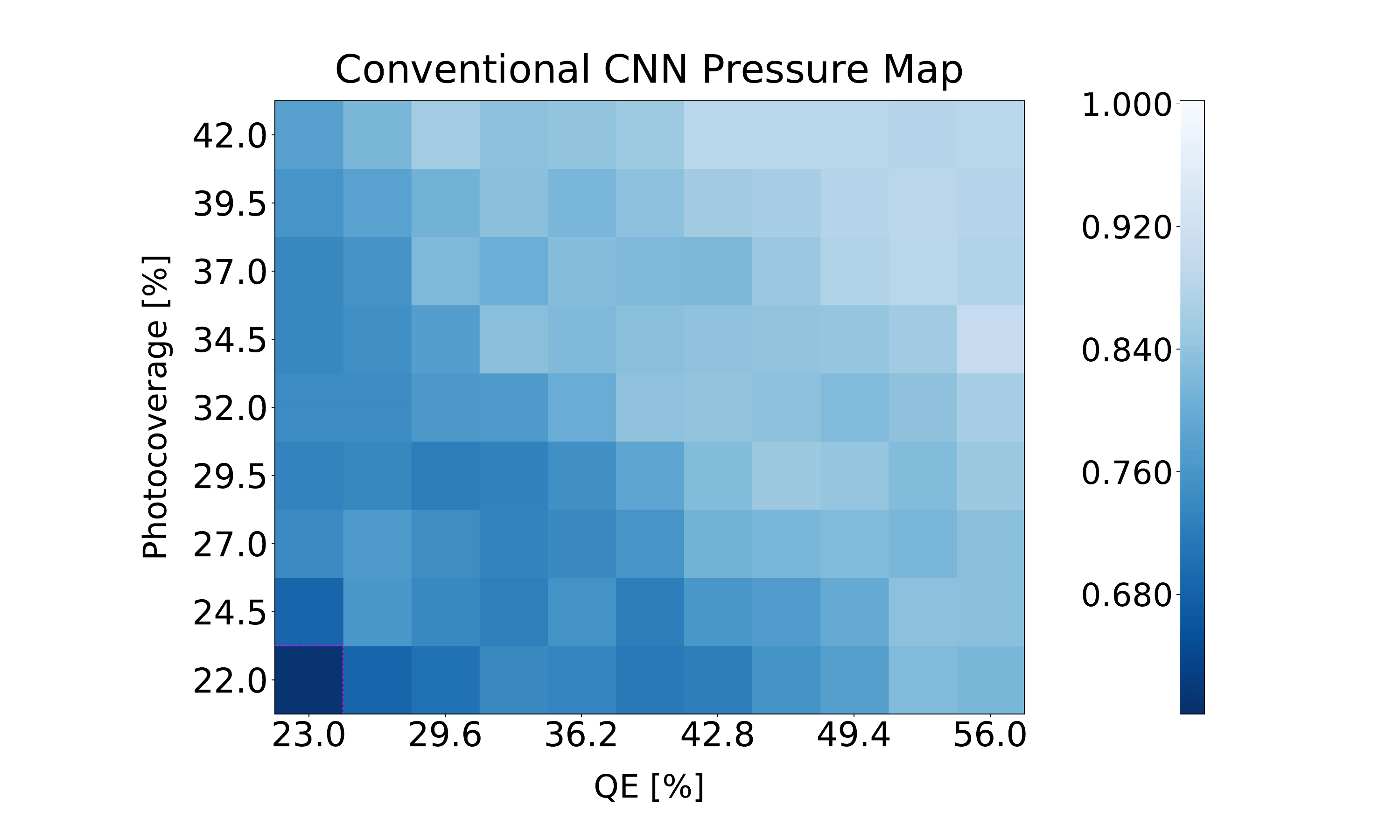}
      \caption{}
      \label{fig:pm_cnn}
    \end{subfigure}
    \begin{subfigure}{1.0\linewidth}
      \centering
      \includegraphics[width=1.0\linewidth,trim={12pc 2pc 8pc 0pc},clip]{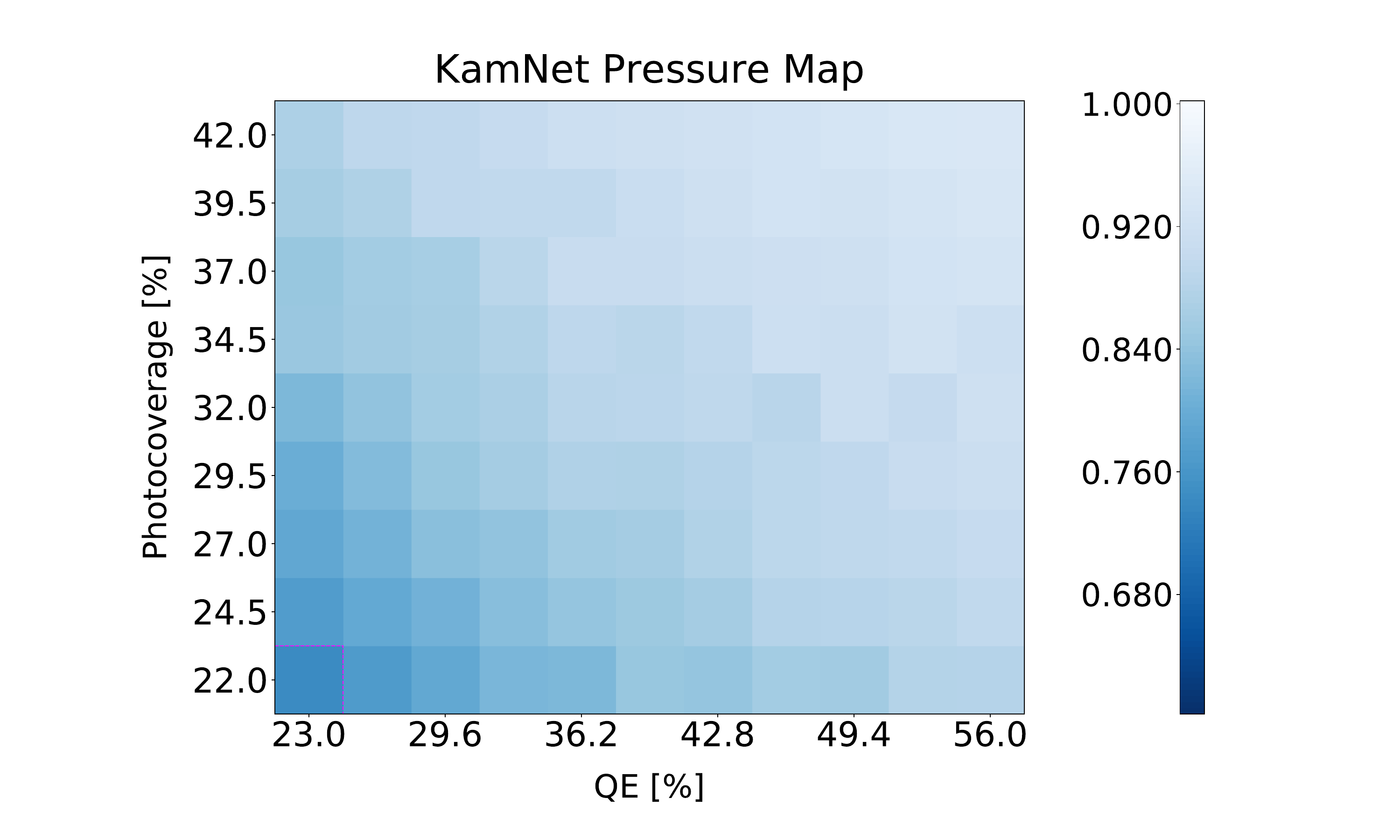}
      \caption{}
      \label{fig:pm_kamnet}
    \end{subfigure}
    \caption{(a) Pressure map produced by conventional CNN described in Reference \cite{nim_paper}. (b) Pressure map produced by KamNet on the same dataset. The values in each cell are the \C\, rejection efficiency at 90\% \Xe\, acceptance. The lower left cell (circled in dashed red) represents the KamLAND–Zen detector configuration.}
   \label{fig:pressure_map}
\end{figure}

In order to provide a direct comparison of the improved performance of KamNet over the conventional CNN, we use the dataset in our previous work~\cite{nim_paper} as a benchmark. As described earlier in Section~\ref{section:simulation}, sim-Fast is used to produce this benchmark dataset and is a simplified version of KamLAND–Zen 400. In order speed up the simulation process, sim-Fast employs a gray-disk model for the PMTs. Both \Xe~ \vbb~ and \C~ events are uniformly distributed within a 3-meter-diameter mini-balloon, which is surrounded by a 13-meter-diameter LS balloon and a 2.5-meter-thick mineral oil buffer volume. Photons are propagated throughout the XeLS mini-balloon, LS balloon, and outer mineral oil buffer volume. Photons that reach the outer boundary of the mineral oil buffer and pass through the gray-disc PMTs are either accepted or rejected based on a uniform quantum efficiency (QE) applied across the gray disk. 

The gray-disc PMT model allows us to vary the detector's QE and the photocoverage, affecting the number of scintillation photons collected. At high QE and the photocoverage, the increased number of scintillation photons provides more information to KamNet, leading to an easier (lower-pressure) classification task; conversely, low QE and photocoverage lead to a more challenging (higher-pressure) classification task. Therefore, we define QE and photocoverage as the two pressure parameters and vary them to generate 99 trials of training datasets. The performance of the deep learning models is demonstrated by training them simultaneously on the 99 trials and displaying the \C~ background rejection efficiencies at 90\% \Xe~ \vbb~ signal acceptance over pressure maps. The resulting pressure maps of the conventional CNN and KamNet are shown in Figure~\ref{fig:pressure_map}.

As seen in Figure~\ref{fig:pressure_map}, KamNet clearly outperforms the conventional CNN over all pressures. In terms of rejection efficiency, KamNet rejects 7.6\% more \C~ events which is obtained by averaging the difference over the entire pressure map. At a photocoverage of 22\% and quantum efficiency of 23\%, which is a similar representation of the real KamLAND–Zen detector configuration, KamNet rejects 74.0\% \C~events compared to 61.5\% for the conventional CNN. Meanwhile, the classification difficulty continuously decreases as we increase photocoverage and quantum efficiency. Therefore, the deep learning models are expected to produce a continuous improvement from the lower-left to the upper-right of the pressure map. KamNet produces a much smoother transition across the pressure map, indicating that KamNet is much more robust to variation of the input data. We believe this stems from KamNet's ability to harness the spherical and spatiotemporal symmetries inherent in the data, while the conventional CNN appears to only be focusing on timing discrepancies. 

\section{\vbb~Decay}
\label{sec:vbb_decay}

\begin{figure}[!htb]
    \centering
    \includegraphics[width=1.0\linewidth,,trim={1pc 0pc 2pc 2pc},clip]{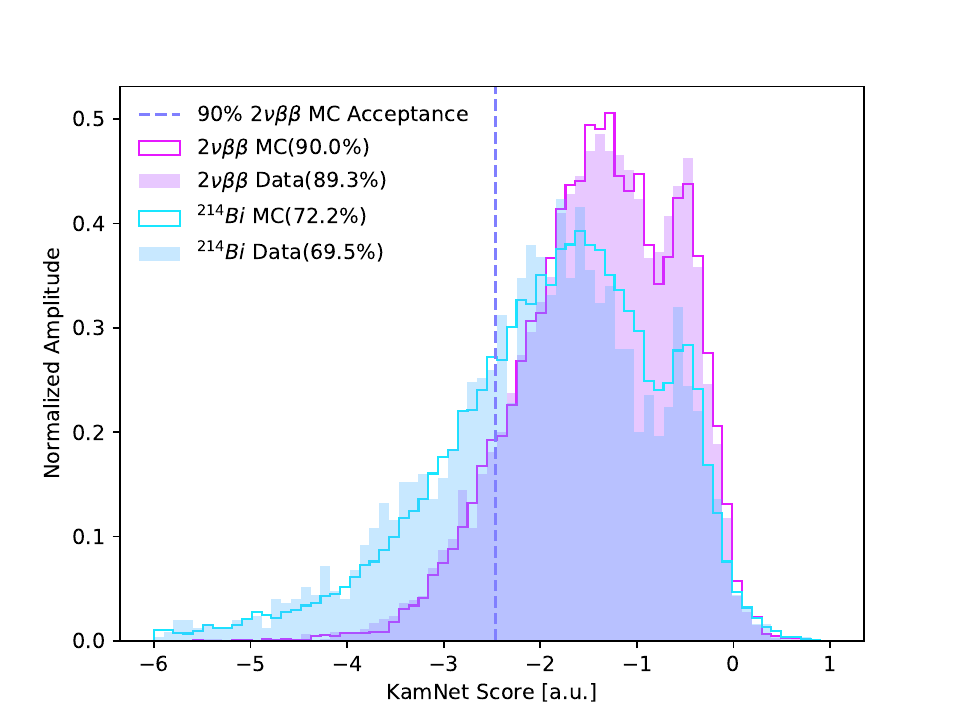}
    \caption{KamNet score spectrum for MC/data of signal events~($^{136}$Xe $2\nu\beta\beta$) and background events~($^{214}$Bi). The number in parentheses shows the survival percentage if we make a cut at 90\% $^{136}$Xe MC acceptance~(purple dashed line).}
    \label{fig:data-mc-agreement}
\end{figure}
We now use the precisely tuned sim-KLZ800 to train KamNet for the \vbb~decay analysis. This study uses events in the energy window 2.0-3.0\,MeV with radii R$<$157\,cm from the center of the XeLS mini-balloon. The signal training set is \vbb~decay to ground state events, and the background training set is $^{214}$Bi events.  We find that the training of KamNet on just one background produces nearly identical classification power compared to a training set formed from a mixture of all the simulated backgrounds. The choice of $^{214}$Bi over other backgrounds is motivated by the fact that it is possible to extract a pure $^{214}$Bi data set from the detector using the prompt-delayed coincidence tagging of $^{214}$Bi-$^{214}$Po decays.

\begin{figure*}[!hbt]
    \begin{subfigure}{0.48\linewidth}
      \centering
      \includegraphics[width=1.0\linewidth,trim={0pc 0 3pc 3pc},clip]{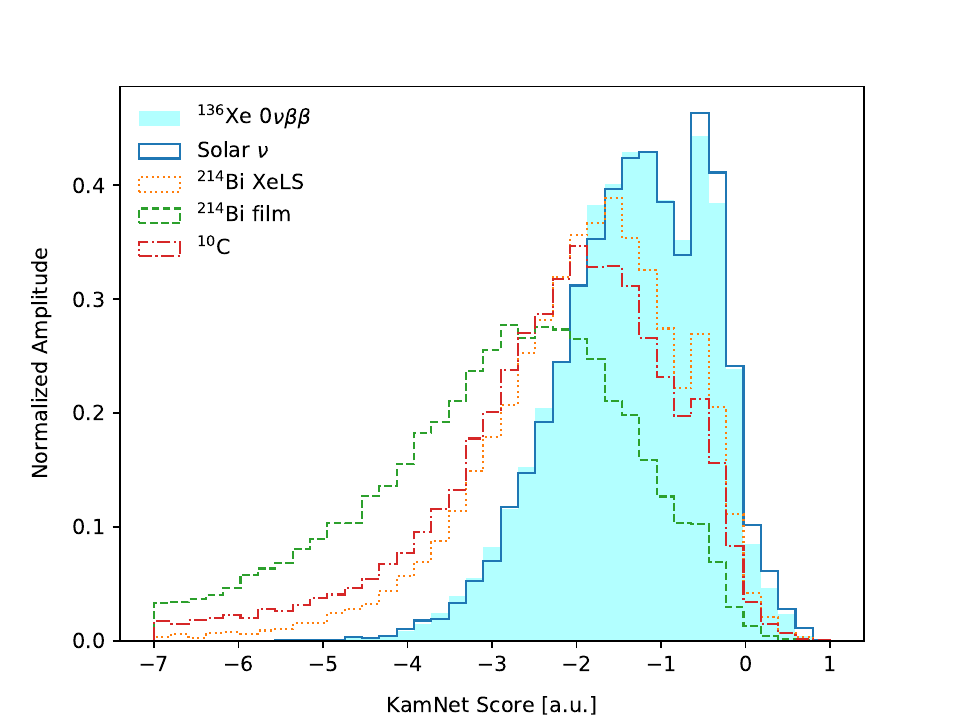}
      \caption{}
      \label{fig:kamnet_vbb_otherbg}
    \end{subfigure}
    \begin{subfigure}{0.48\linewidth}
      \centering
      \includegraphics[width=1.0\linewidth,trim={0pc 0 3pc 3pc},clip]{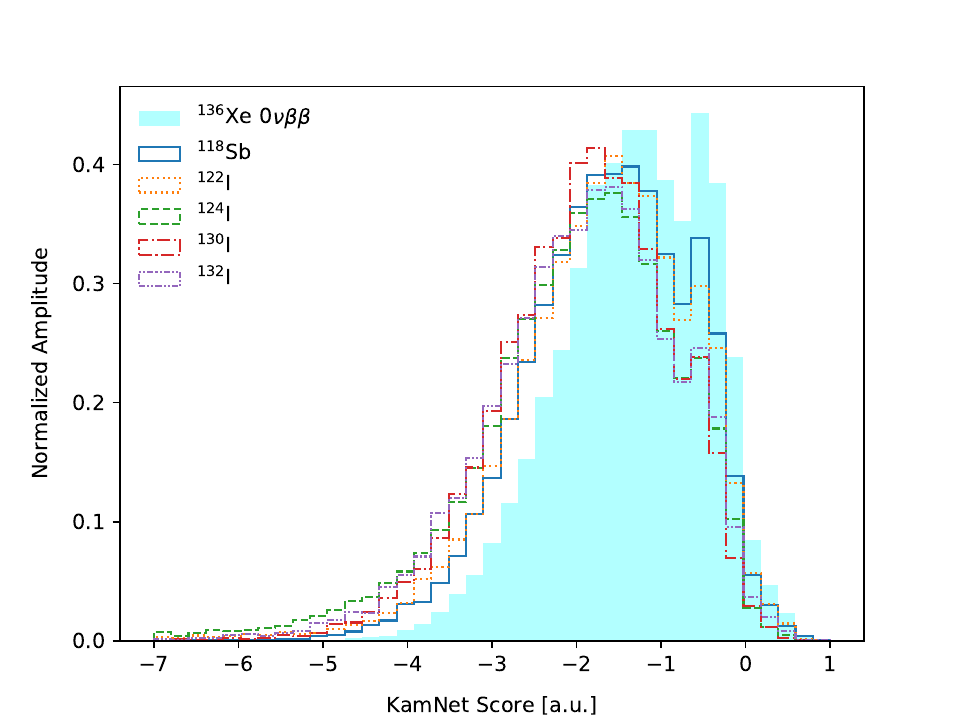}
      \caption{}
      \label{fig:kamnet_vbb_llbg}
    \end{subfigure}
    \caption{(a) KamNet score spectrum for common backgrounds in KamLAND–Zen 800, including solar neutrino, \Bi, and $^{10}$C backgrounds. (b) KamNet score spectrum for dominant long-lived spallation backgrounds in energy ROI.  All histograms have been normalized to unity. Except for Solar $\nu$, all backgrounds has lower KamNet score compared to \Xe, thus they can be efficiently rejected by making cut on KamNet score.}
    \label{fig:kamnet_vbb}
\end{figure*}

To check the MC-data agreement, we study signal-like \vvbb~events and background-like events in data and MC. For the signal-like \vvbb~events, we select events from 1.7\,MeV to 2.2\,MeV within the 157\,cm radius, a region dominated by \vvbb~events. For the background-like events, we use $^{214}$Bi-$^{214}$Po tagged events from 2.0\,MeV to 3.0\,MeV within the 157\,cm radius. The selection criteria are identically applied to both data and MC. KamNet is then applied to the MC and data events. The resulting KamNet scores are shown in Figure~\ref{fig:data-mc-agreement}. Excellent agreement is observed between the data and MC spectrum shape. 

To quantify the MC-data agreement, we define a cut at the KamNet score corresponding to 90\% \vvbb-MC acceptance. At this same KamNet score, the corresponding \vvbb  data has a slightly smaller acceptance of 89.3\% due backgrounds in the data. This is consistent with the results in this region presented in Ref.~\cite{KamPRL}. The acceptance of \Bi~data events are 69.5\% compared to 72.2\% of \Bi~MC events. \Bi~data events are rejected more efficiently because 14.1\% of the events originated on the balloon film. Film \Bi~are rejected much more efficiently by KamNet, as shown in Table~\ref{tab:kamnet_vbb_result}. Adding 14.1\% of film \Bi~into the \Bi~MC will reduce the data-MC difference to only 1.7\%. These effects are carefully quantified and included in the upstream data analysis~\cite{KamPRL} as systematic uncertainties.


\subsection{Network Interpretability}
\label{subsec:interpret}

The rejection power of KamNet comes from distinguishing strictly single-vertex events from closely spaced multi-vertex events such as $\beta^{\pm}$ decay with $\gamma$ cascade. This is demonstrated in Figure~\ref{fig:kamnet_vbb_otherbg} using sim-KLZ800 by comparing the KamNet score spectrum of the $\gamma$ cascade decays like \Bi~or $^{10}$C with the $\beta$ events produced by the elastic scattering of solar neutrinos. We again set a cut on the KamNet score corresponding to the acceptance of 90\% of the \vbb~signal. Using this criterion, KamNet rejects 27.1\% of \Bi~events from within the XeLS and 59.6\% of \Bi~events from the mini-balloon film. In comparison, the rejection of solar neutrino electron elastic scattering events is only 9.8\%. 

We can extend this study to long-lived spallation products, unstable light isotopes produced by high energy cosmic muons interacting in the liquid scintillator. Each long-lived spallation isotope undergoes a $\gamma$ cascade with a unique relative intensity. It turns out that the isotope with highest relative intensity~($^{90}$Nb) has the highest rejection efficiency (35.1\%), while the isotope with lowest relative intensity~($^{137}$Xe) has the lowest rejection efficiency (12.1\%). Aggregating all long-lived isotopes, we obtain a Pearson correlation coefficient of 0.49 between rejection efficiency and the relative intensity of the $\gamma$ cascade. This indicates that the rejection power of KamNet is moderately correlated with the $\gamma$ cascade. In near future, we will conduct a systematic study to comprehensively understand how the $\gamma$ cascade affect KamNet's performance.
\begin{figure*}[hbt!]
    \begin{subfigure}{0.49\linewidth}
      \centering
      \includegraphics[width=1.0\linewidth,trim={3pc 0 4pc 0pc},clip]{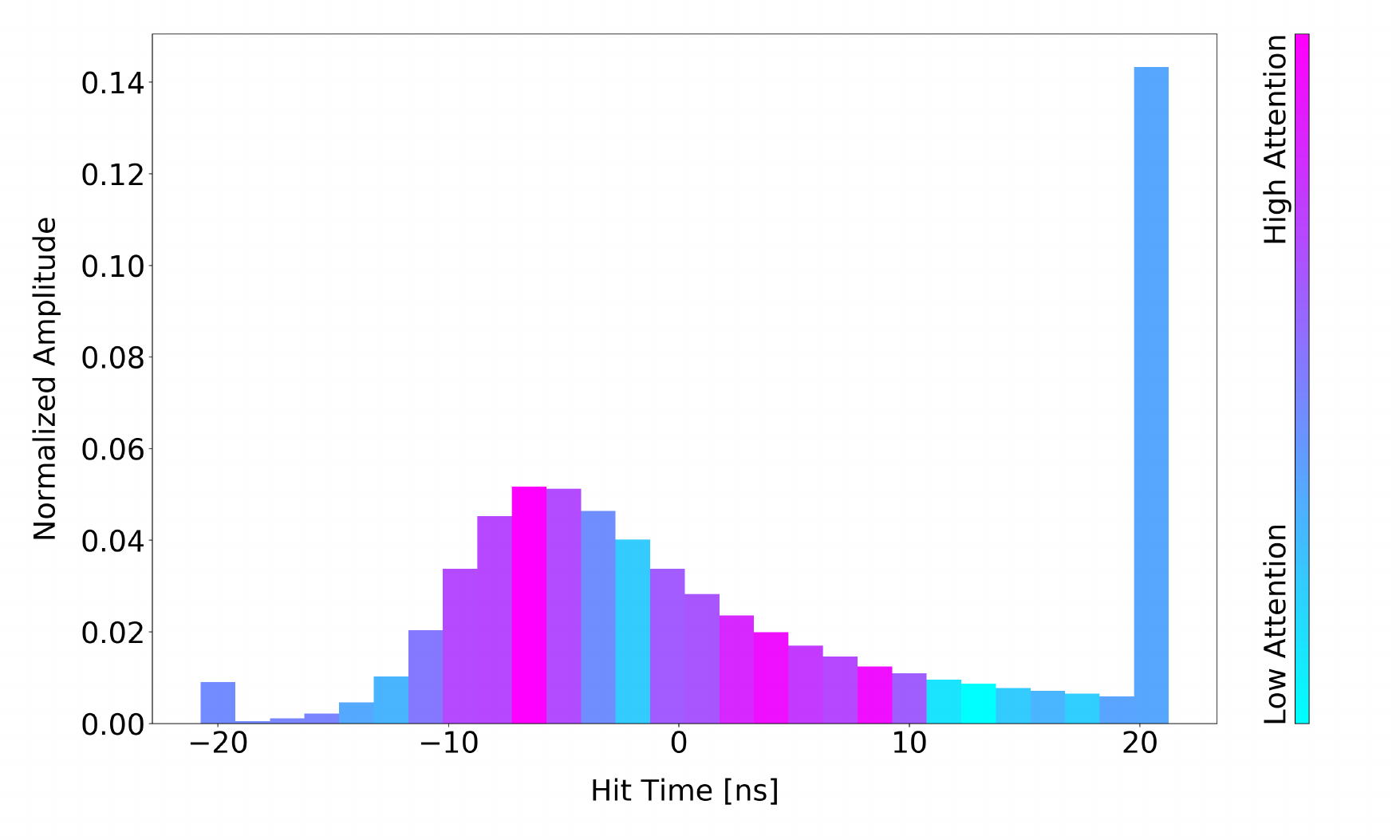}
      \caption{}
      \label{fig:kamnet_attention_bi214}
    \end{subfigure}
    \begin{subfigure}{0.49\linewidth}
      \centering
      \includegraphics[width=1.0\linewidth,trim={3pc 0pc 4pc 0pc},clip]{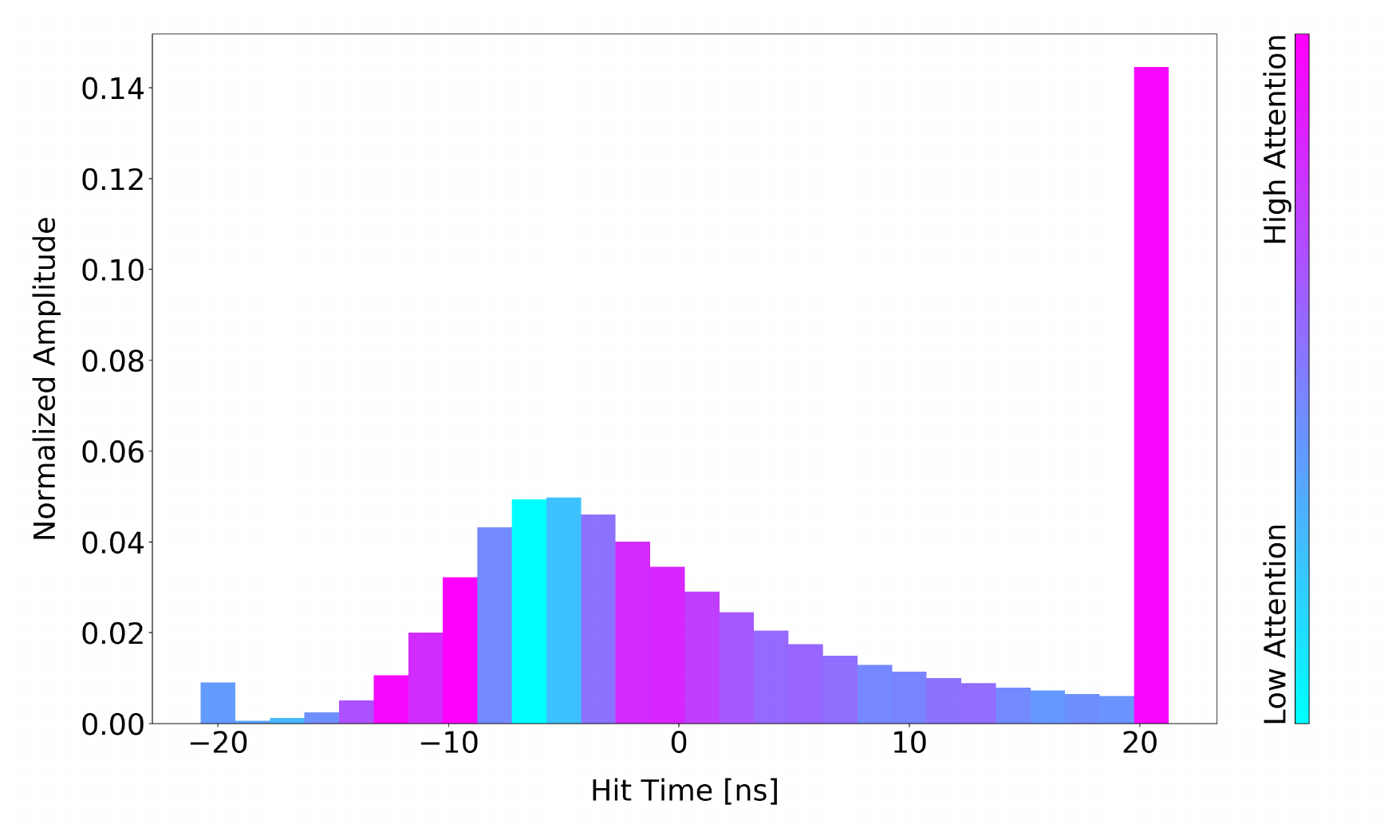}  
      \caption{}
      \label{fig:kamnet_attention_c10}
    \end{subfigure}
    \caption{The attention score plot of KamNet. The shape of plot shows the scintillation time profile of KamLAND–Zen events, and the color shows the relative attention score of each time slice. KamNet relies heavily on the high attention (magenta) region to make classification decisions, not as much on the low attention (cyan) region. In plot (a), KamNet is trained to reject \Bi~backgrounds from \vbb~signal. In plot (b), KamNet is trained to reject $^{10}$C~ backgrounds from \vbb~signal.}
    \label{fig:kamnet_attention}
\end{figure*}

The identification of closely spaced multi-vertex events can be directly visualized using the attention score in Equation~\ref{eqn:attention_score}. The attention score is a probabilistic float point number assigned to each slice of the spatiotemporal hit map. KamNet relies heavily on time slices with high attention score to make classification decisions, not as much on slices with low attention score. 

Figure~\ref{fig:kamnet_attention} shows the relative attention score for KamNet trained to reject different backgrounds. The shape of plot shows the scintillation time profile of KamLAND–Zen events. The bin width are 1.5\,ns, identical to the input series of spherical hit maps. The first and last bins are populated since it contains overflow and underflow hits. The color of each bin indicates the relative attention score on this time slices as assigned by KamNet. In Figure~\ref{fig:kamnet_attention_bi214}, KamNet is trained to reject \Bi~backgrounds, thus most of the attention is placed between 5-10\,ns when the $\gamma$ casacade occurs. The second highest attention is placed near the scintillation peak. On the other hand, KamNet in Figure~\ref{fig:kamnet_attention_c10} is trained to reject $^{10}$C backgrounds. In this case, the highest attention is placed at the rising edge, since $^{10}$C undergoes $\beta^{+}$ decay to produce a pair of $\gamma$. These $\gamma$s will Compton scatter and deposit their energy on a slightly longer timescale than $\beta^{-}$ decay. In some case, the $e^{+}$ and $e^{-}$ will form a positronium which further delay the energy deposition and adds on to the effect of rising edge. This effect was also observed in our previous work~\cite{nim_paper}. Besides the positronium effect, a 718\,keV $\gamma$ is also released in 0.7\,ns after \C~decays to $^{10}$B excited states. KamNet also pay attention to the falling edge of scintillation time profile to capture this $\gamma$ casacade.

The biggest difference between Figure~\ref{fig:kamnet_attention_bi214} and Figure~\ref{fig:kamnet_attention_c10} is the last overflow bin. In Figure~\ref{fig:kamnet_attention_bi214}, the last bin contains secondary effects such as absorption, re-emission, scattering and dark noise. These effects are identical in both signal and background events, thus KamNet does not to pay attention to the last bin. On the contrary, KamNet places a significant amount of attention to the overflow bin in Figure~\ref{fig:kamnet_attention_c10}. KamNet in Figure~\ref{fig:kamnet_attention_c10} is trained with $^{10}$C with ortho-positronium half-life. Ortho-positronium possesses a lifetime of 3\,ns in Liquid Scintillator~\cite{betapm}. Along with the delay caused by LS timing response, ortho-positronium allows the decay to delay and leak physical information into the last overflow bin, and KamNet is able to extract that information to make classification decision.

\subsection{Data Cleaning}
\label{subsec:cleaning}

With the ability to interpret the network, we can now use it to identify periods of high background rates. In KamLAND, we observe that detector operating conditions, such as temperature fluctuations in the buffer liquids surrounding the PMTs or vibrations from construction on the deck above the detector, can lead to periods of increased convection inside the inner detector volumes. Convection in the mini-balloon tends to pull contaminants off of the balloon surface and into the main $0\nu\beta\beta$ analysis volume. 
These contaminants are primarily closely spaced multi-vertex events KamNet can efficiently identify. Therefore, likelihood profiles of KamNet score are constructed based on the MC simulation of several representative isotopes, including both $\beta$-like signal and $\beta^{\pm}+\gamma$ backgrounds. The likelihood profiles are then fitted to data to extrapolate periods with high background concentration. 
This is a powerful tool that is used in conjunction with other monitoring tools and logbooks to veto these periods of data instability as described in Ref.~\cite{klz800_prl}.

\subsection{Background Rejection}
\label{subsec:background}

\renewcommand*{\arraystretch}{1.4}
\begin{table}[!h]
\begin{tabular}{|c|ccc|}
\hline
Isotopes&Type & ROC AUC & Rejection\\
\hline
\vbb & Signal & 0.5 & 10\% \\
Solar $\nu$ & $e^{-}$ & 0.49&9.5\% \\
$^{10}$C & $\beta^{+}+\gamma$ & 0.72&40.0\% \\
\Bi~XeLS & $\beta^{-}+\gamma$ & 0.65&27.0\% \\ 
\Bi~film & $\beta^{-}+\gamma$ & 0.83&58.8\% \\
$^{118}$Sb& $\beta^{+}+\gamma$, LL & 0.59&18.3\%\\
$^{122}$I& $\beta^{+}+\gamma$, LL & 0.61&22.2\%\\
$^{124}$I& $\beta^{+}+\gamma$, LL & 0.67&30.6\%\\
$^{130}$I& $\beta^{-}+\gamma$, LL & 0.67&27.2\%\\
$^{132}$I& $\beta^{-}+\gamma$, LL & 0.66&28.5\%\\
\hline
\end{tabular}
\caption{ Result of trained KamNet classifier on \vbb~analysis. The second column of the table indicates the type of decay the isotope undergoes. $e^{-}$ indicates a strictly single vertex, $\beta$ like events. $\beta^{\pm}+\gamma$ indicates a $\beta$ decay with $\gamma$ casacade, and LL stands for long lived spallation backgrounds.  The KamNet rejection percentage is evaluated at 90\% signal acceptance (or equivalence speaking, 10\% signal rejection). ROC AUC is the area under curve of Receiver Operating Characteristic curve\cite{ROC_paper}, higher AUC under ROC curve indicates the better separation between signal and background.} 
\label{tab:kamnet_vbb_result}
\end{table}


On average, KamNet rejects 27\% of internal backgrounds and 59\% of film backgrounds. Furthermore, the background rejection of KamNet doesn't rely on high-level analysis cuts, like prompt-delayed coincidence tagging, or hardware upgrades. Thus, the background rejection factor has a multiplicative effect when applied to any standard physics analysis. Since long-lived spallation products are the major sources of background in the KamLAND–Zen region of interest, and an efficient coincidence tag is challenging because of their long half-life, KamNet plays a key role in pushing the KamLAND–Zen \vbb~limit forward. To evaluate the performance of KamNet, we selected five abundant long-lived spallation backgrounds within the KamLAND–Zen ROI.  The rejection efficiencies of those backgrounds are listed in Table~\ref{tab:kamnet_vbb_result}.
We now use KamNet's rejection power against XeLS backgrounds to estimate the expected sensitivity boost for the \vbb~search in KamLAND–Zen. To keep the calculation simple, a counting experiment model is used to estimate the sensitivity. This estimate is conservative compared to the fitting of the energy spectrum performed in a full-scale KamLAND–Zen analysis. In this model, KamLAND–Zen's sensitivity is proportional to $S/\sqrt{B}$, where S is the number of signal events and B is the number of background events in the ROI. If a classifier rejects 30\% of the background while preserving 90\% of the signal, then the sensitivity will be boosted by:
\begin{equation}
    T_{0\nu}^{1/2} \propto \frac{90\%S}{\sqrt{(100\%-30\%)B}} = 1.076\frac{S}{\sqrt{B}}
    \label{eqn:sensitivity_factor}
\end{equation}
Here, 90\% is the true positive rate (TPR) and 30\% is the true negative rate (TNR). One minus the true negative rate is the false positive rate (FPR), indicating the percentage of backgrounds remaining after cut. The number 1.076 is the sensitivity factor $\mathcal{S}$ for this classifier, corresponding to a 7.6\% increase in sensitivity. In KamLAND–Zen, we have more than one type of background in the ROI, so we have to take all of them into account. From preliminary fitting results~\cite{KLZ_prelim}, 58.2\% of backgrounds are long-lived spallation backgrounds which can be efficiently rejected by KamNet, and 41.8\% of backgrounds are irreducible \vvbb-like backgrounds.  Therefore, we can estimate the sensitivity factor $\mathcal{S}$ with the following equation:
\begin{equation}
    \mathcal{S} =\frac{\mathrm{TPR}}{\sqrt{\mathrm{FPR}_{LL} \cdot 58.2\% + \mathrm{FPR}_{2\nu} \cdot 41.8\%}}
    \label{eqn:sensitivity_XeLS}
\end{equation}
Since KamNet does not have any rejection power against \vvbb-like backgrounds, FPR$_{2\nu}$ is equivalent to TPR. $\mathrm{TPR}$ and $\mathrm{FPR}_{LL}$ are obtained by making cut at a given KamNet cutting threshold and evaluating on sim-KLZ800. Based on this calculation, we are able to evaluate the sensitivity factor on all possible cutting thresholds. While accepting 90\% signal events, KamNet boost the \vbb~search sensitivity by 2.2\%. This boost can be further enhanced by optimizing the energy selection to reduce the amount of \vvbb~backgrounds.

The 2.2\% boost is extremely conservative. The KamLAND-Zen 800 analysis is a multi-dimensional fit in energy and position. KamNet's ability to reject backgrounds with peak-like features in the ROI, amplifies its power in the region of interest providing much larger gains in sensitivity. For this reason, our future work focuses on a native propagation of the output of KamNet to the Bayesian extraction of the \vbb~result. This includes a modification of KamNet to produce joint Bayesian priors for each fitting spectrum and then integrate them into the Bayesian analysis.

%

\subsection{Fiducial Volume Expansion}
\label{subsec:fv}
\begin{figure}[htb!]
    \centering
    \includegraphics[width=1.0\linewidth,trim={2pc 0 0 5pc},clip]{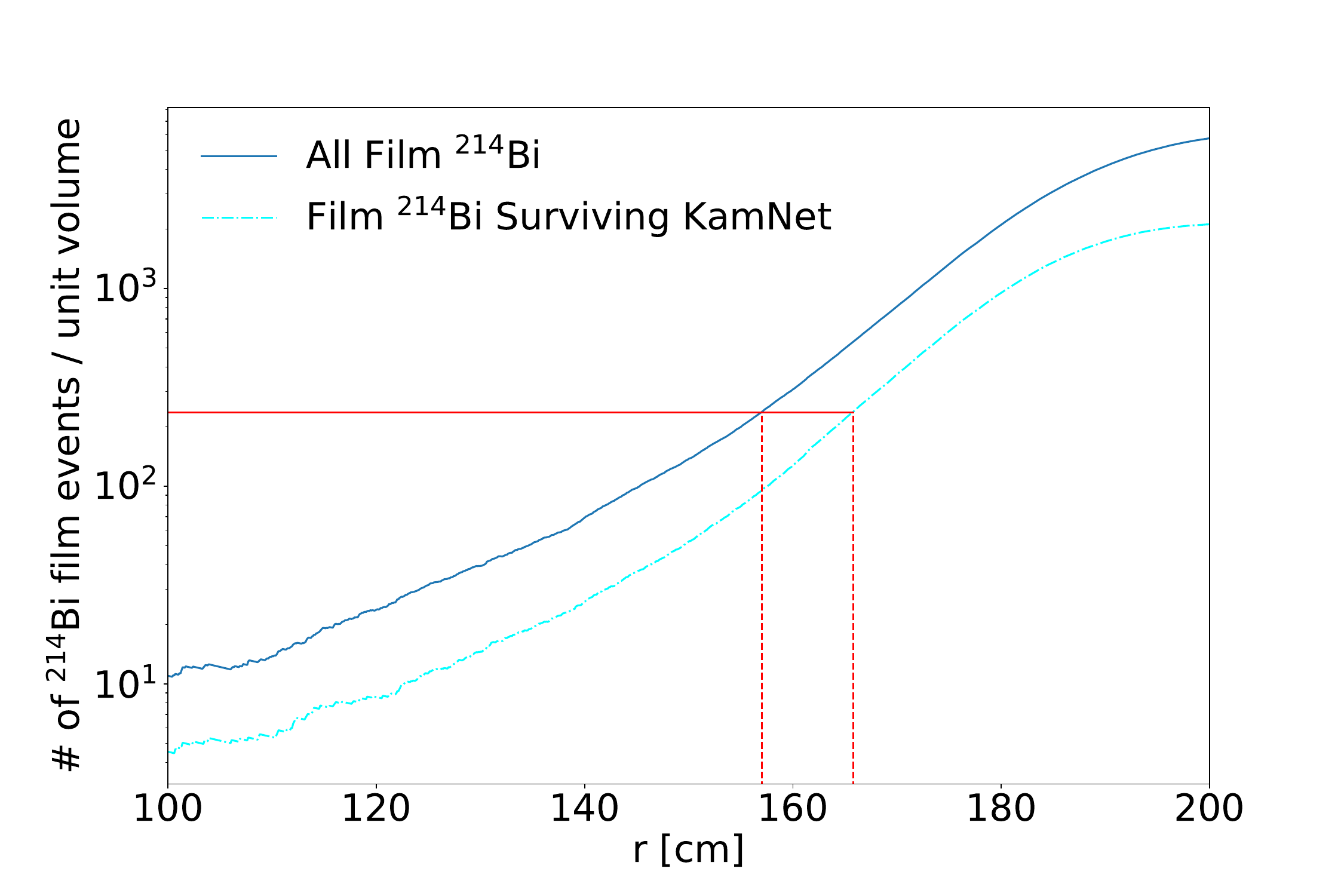}
    \caption{Result of KamNet fiducial volume study based on sim-KLZ800. The dark blue curve shows the film background rate before KamNet cut, and the cyan dashed curve shows film background rate after KamNet cut.}
    \label{fig:sensitivity_film}
\end{figure}

KamNet's rejection power against film backgrounds can also lead to an independent sensitivity boost. According to Table~\ref{tab:kamnet_vbb_result}, KamNet efficiently rejects 59\% of film background while maintaining 90\% signal acceptance. Since film background is the major limitation on KamLAND–Zen fiducial volume, KamNet enable us to expand fiducial volume to gain more exposure. Figure~\ref{fig:sensitivity_film} shows the film background rate before and after KamNet cut is applied. Based on this figure, KamNet allows us to expand fiducial volume from $R<157$\,cm to $R<165.8$\,cm without increasing the film background level. This corresponds to a 17.7\% increase in fiducial volume and exposure. In monolithic detectors, the sensitivity is proportional to the exposure in the following way:
\begin{equation}
    T^{1/2}_{0\nu} \propto \sqrt{\frac{\alpha\epsilon M}{B\delta E}}
    \label{eqn:sensitivity_monolithic}
\end{equation}
where $\alpha$ is the isotopic abundance, $\epsilon$ is the detection efficiency, M is the exposure, B is the number of backgrounds and $\delta E$ is the energy resolution. Using this equation, we quote a 8.5\% sensitivity boost from 17.7\% increase in exposure. Once again, this is a conservative estimate.

The two sensitivity boosts we discussed above are mutually independent. Even within the expanded fiducial volume, KamNet can still efficiently reject XeLS background events. Therefore, the final sensitivity boost is quoted by multiplying 2.2\% XeLS sensitivity boost to 8.5\% film sensitivity boost, resulting in a 10.8\% overall sensitivity boost. With the help of KamNet, KamLAND–Zen unleashes its full detection power toward \vbb~decay.

\section{\vvbb~Decay to Excited States Dataset}
\label{sec:excitedstates}

\begin{figure}[htb!]
    \centering
    \includegraphics[width=1.0\linewidth,trim={0 5pc 0 2pc}]{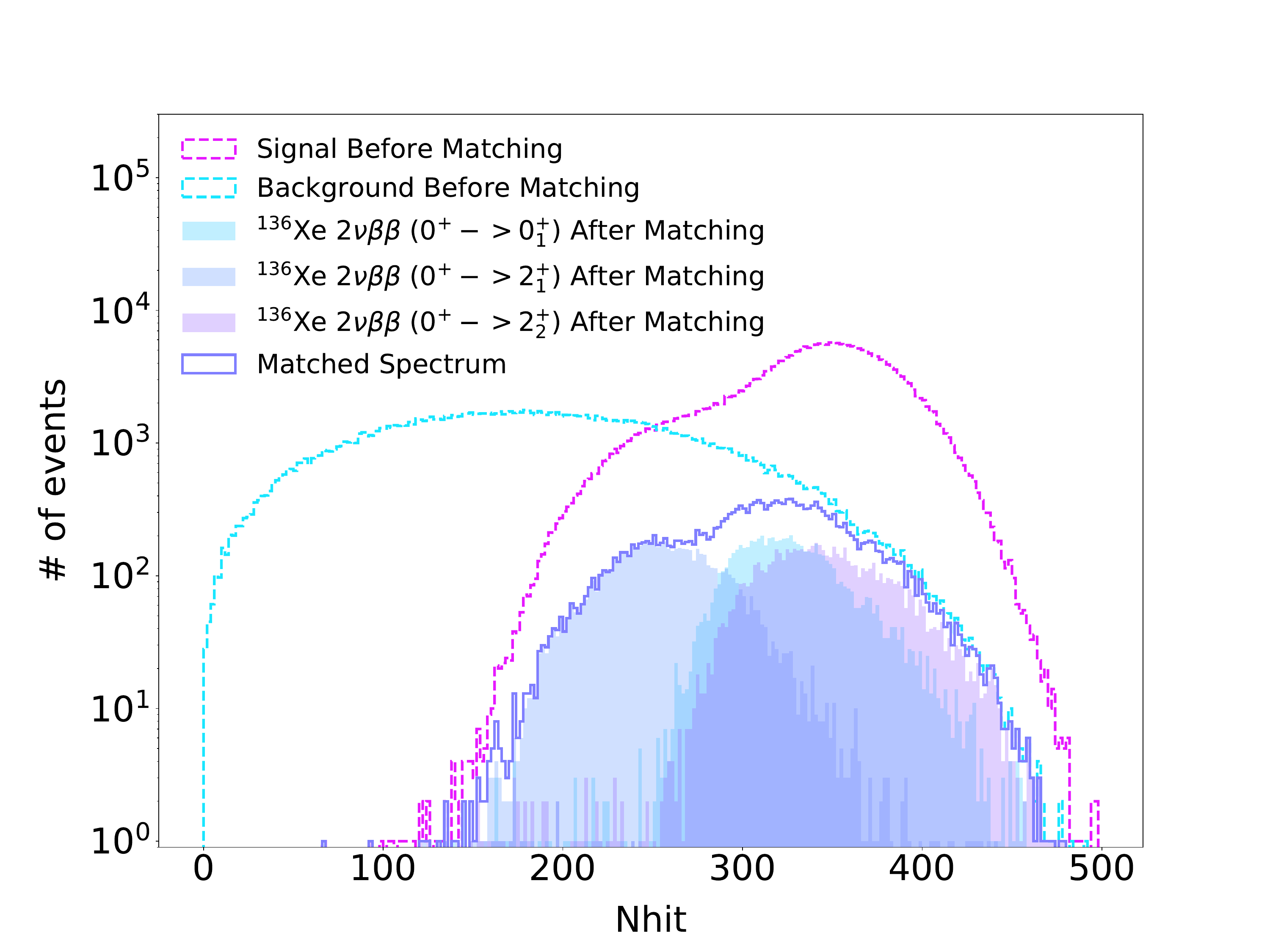}
    \caption{Effect of Nhit matching on the signal and background Nhit distributions. The signal distribution is shown being equally divided into the three excited states.}
    \label{fig:nhit_matching}
\end{figure}

\begin{figure*}[hbt!]
    \begin{subfigure}{0.49\linewidth}
      \centering
      \includegraphics[width=1.0\linewidth,trim={5pc 0pc 0 6pc},clip]{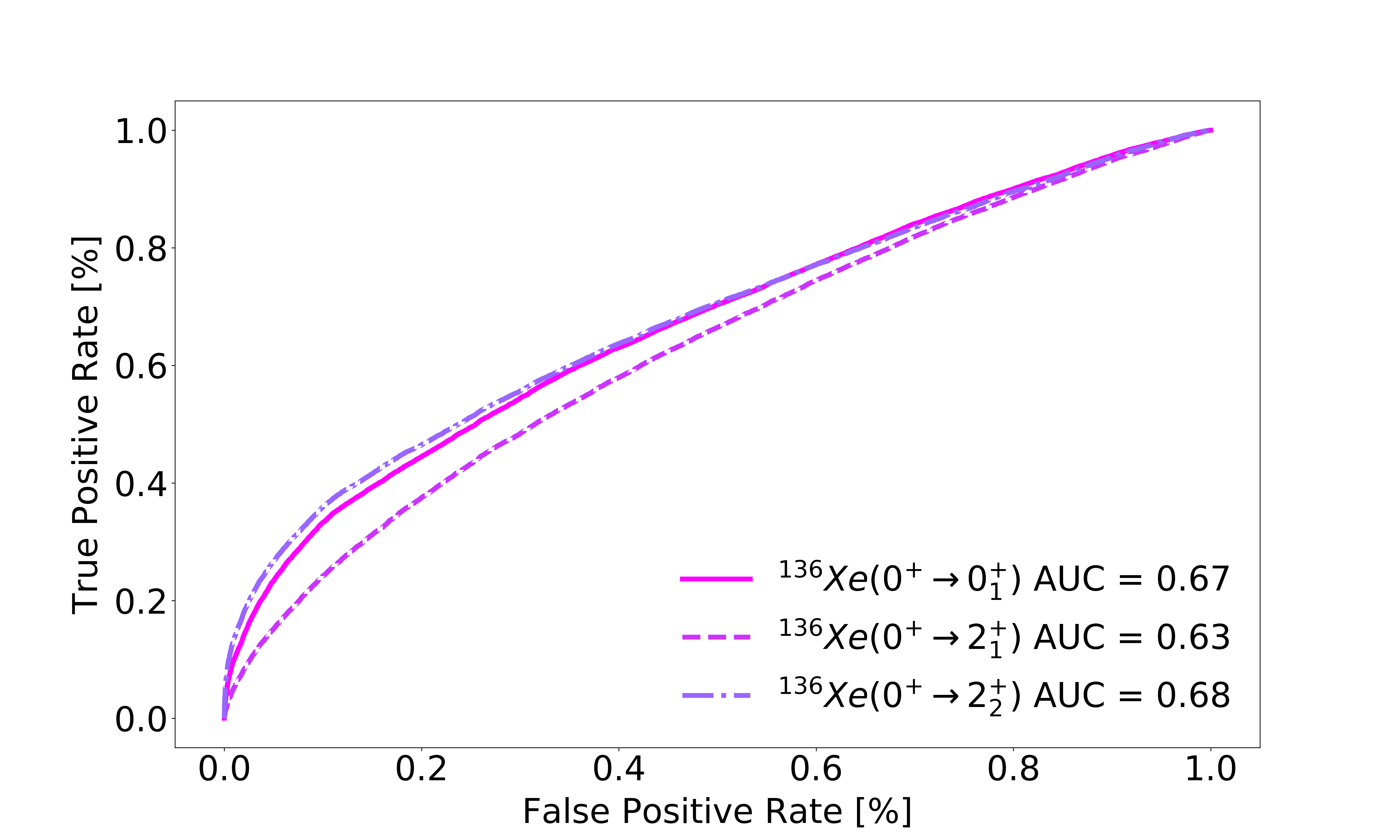}
      \caption{}
      \label{fig:des_roc}
    \end{subfigure}
    \begin{subfigure}{0.49\linewidth}
      \centering
      \includegraphics[width=1.0\linewidth,trim={5pc 0pc 0 6pc},clip]{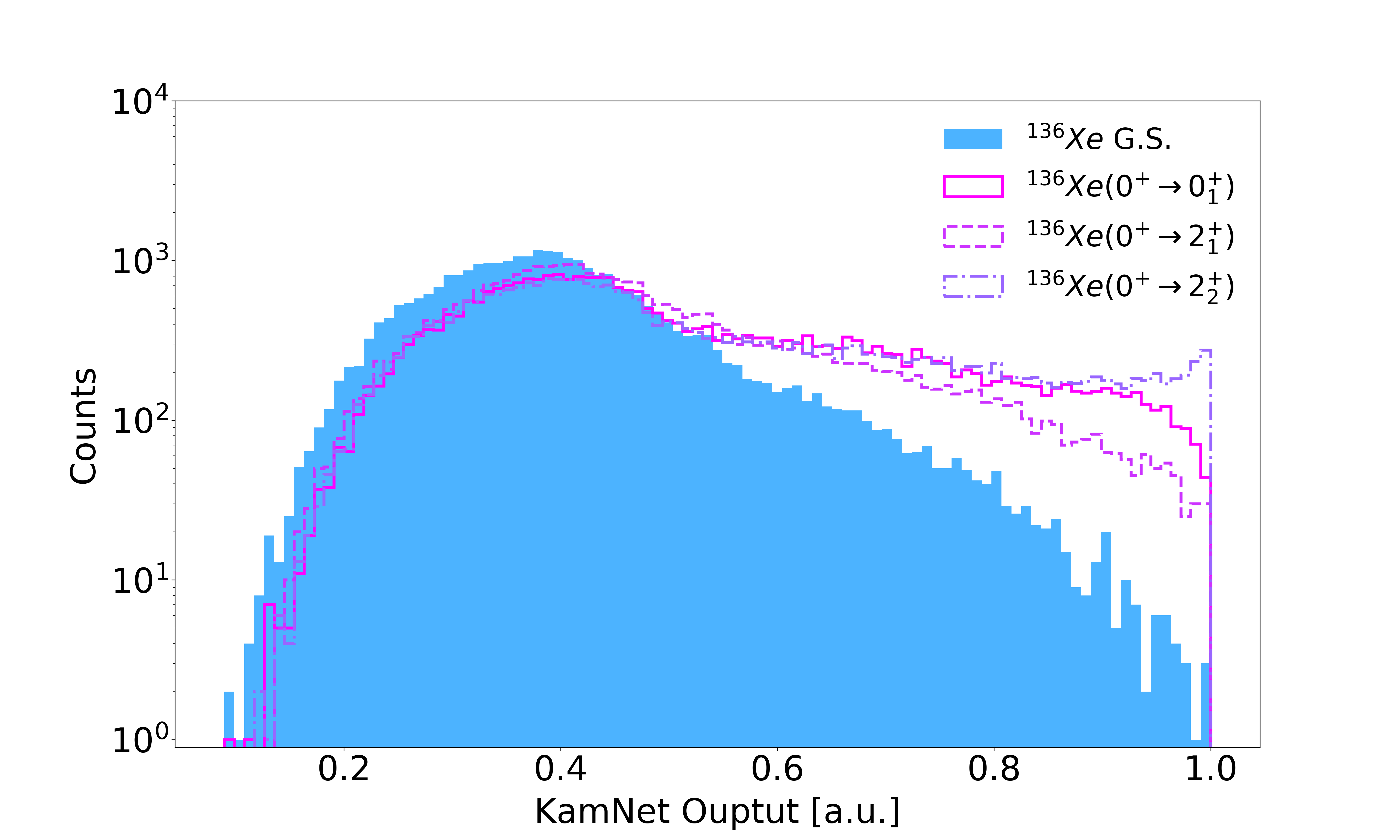}
      \caption{}
      \label{fig:des_distribution}
    \end{subfigure}
    \caption{(a) The ROC curve of KamNet output on each \Xe~decay to excited state signal vs. \Xe~ ground state backgrounds. (b) KamNet score spectrum of 3 \Xe~decay to excited states and \Xe~ ground state.}
\end{figure*}
\begin{figure*}[hbt!]
    \centering
    \includegraphics[width=0.95\linewidth,trim={9pc 5pc 10pc 9pc},clip]{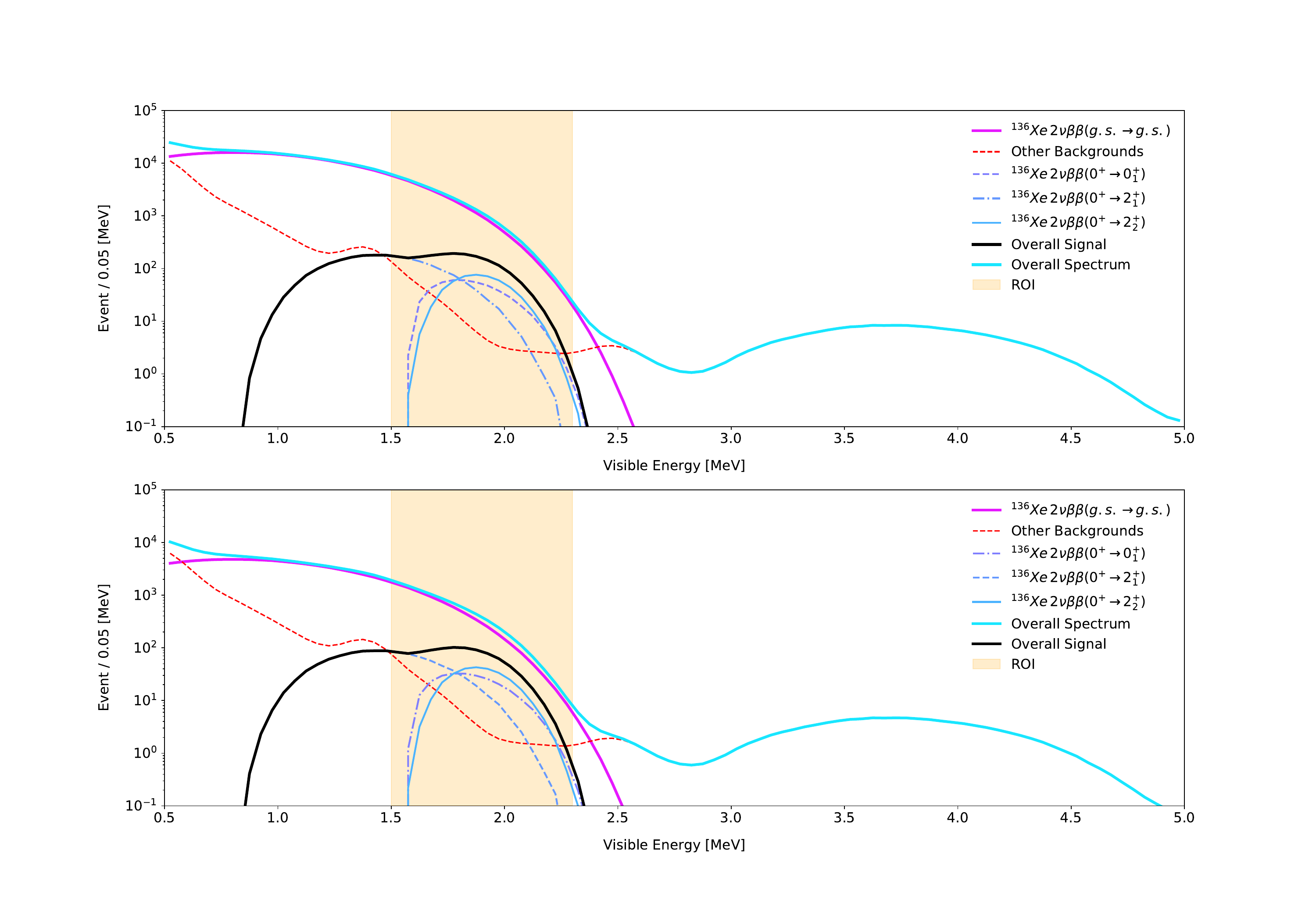}
    \caption{The KamLAND–Zen 800 MC energy spectrum before applying KamNet (top) and after applying KamNet (bottom). The ratio between \vvbb~DES and \vvbb~DGS is obtained from Reference~\cite{kamland_des}. The DES distribution corresponds to 90\% confidence level upper limit.}
    \label{fig:kamnet_effect}
\end{figure*}

In order to quantify KamNet's performance on \vvbb~decay of \Xe~to excited states, we use the sim-RAT to generate training and validation datasets. The three main excited-state decays are simulated and aggregated together into one signal, and the decay to the ground state is simulated as the background. The dataset containing both signal and background events is split into training and validation datasets with a 7:3 ratio. KamNet is trained on the training dataset to produce a binary classifier and its performance is evaluated separately for each type of excited-state decay against the decay to the ground state.

Throughout this study we refer to the number of non-zero cells in the spatiotemporal hit map as `Nhit' where the total number of cells is equal to 40,432. Energy deposits in XeLS of about 2.5 MeV produce roughly 350 Nhits on average. Early studies showed that KamNet would attempt to classify events only by Nhit and ignore all other information when Nhit is very different between the signal and background events. Therefore, we perform something called `Nhit matching' prior to training.  This means that both signal and background events are sampled from an Nhit distribution which is formed from the overlap between the signal and background Nhit distributions.  During Nhit matching the signal Nhit distribution is divided equally among the three different excited state decays. The matching loops through all possible Nhits and, at each step, randomly samples the same number of events from signal and background datasets without replacement. The signal and background Nhit distributions before, and after, Nhit matching are shown in Figure~\ref{fig:nhit_matching}. A total of 300,000 background events and 100,000 signal events forms the raw training dataset. After matching, the background and signal distributions each contain 33,498 events, and the events in the signal distribution are divided equally among the three excited states. 

After pre-processing, the dataset is fed into KamNet for training in PyTorch~\cite{pytorch}. KamNet is trained over 30 epochs and the binary cross-entropy loss is minimized with the ADAM optimizer~\cite{Adam}. After training, the result is evaluated in the form of Receiver Operating Characteristic (ROC) curves. The ROC curve of each excited state signal is plotted in Figure~\ref{fig:des_roc}, where a higher area under curve (AUC) indicates better distinguishability between signal and background events. If the threshold is set to reject 70\% of the background, the signal acceptance efficiency $\epsilon$ for $(0^+ \rightarrow 0^+_1)$, $(0^+ \rightarrow 2^+_1)$, and $(0^+ \rightarrow 2^+_2)$~events are 54\%, 49\%, and 56\%, respectively. Based on the decay half-lives provided by \cite{des_branching}, we estimated the branching ratio $r$ of $(0^+ \rightarrow 0^+_1)$, $(0^+ \rightarrow 2^+_1)$, and $(0^+ \rightarrow 2^+_2)$ to be 0.2970, 0.002366 and $3.246\times10^{8}$ at $g_{A}$ of 0.60. Thus the overall signal efficiency is calculated using:
\begin{equation}
    \epsilon_{DES} = \frac{\sum_{i}r_{i}\epsilon_{i}}{\sum_{i}r_{i}}
    \label{eqn:signal_eff_DES}
\end{equation}
The summation is conducted over 3 excited states decay and the final $\epsilon_{DES}$ is 54.0\%.

To evaluate the impact of KamNet's background rejection power on the excited state analysis, we use a recent estimate of the backgrounds in KamLAND–Zen 800~\cite{aobo_thesis}. The possible improvement on the excited state analysis after applying KamNet is illustrated in Figure~\ref{fig:kamnet_effect}. The spectrum amplitudes of the excited states are fixed using the lower half-life limits (at 90\% C.L.) from an earlier analysis with KamLAND–Zen 400~\cite{kamland_des}. The impact of KamNet on the excited-state analysis is estimated from the improvement in the signal-to-noise ratio (S/N). It is calculated using the sum of all three signals inside an energy window of 1.5-2.3\,MeV. In this region, the excited state decay signals are closely spaced multi-vertex events with $\gamma$ cascades. The dominant background in this region strictly single-vertex, \vvbb~decay to the ground state, which constitutes roughly 99\% of the background. In addition, backgrounds such as $^{214}$Bi, $^{11}$C, $^{122}$I, $^{124}$I, $^{130}$I, and $^{118}$Sb could also leak into this energy window. Unlike the situation in Section~\ref{sec:vbb_decay}, these additional backgrounds contain $\gamma$ cascades and look more similar to the signal than to the dominant background. Due to limited time and computing resources, we chose to only use KamNet to reject the \vvbb~decay to ground state and not run it over all of the additional backgrounds. Instead, we conservatively assume that KamNet will treat these backgrounds the same way it treats the excited state signals. Therefore, we apply an acceptance factor of 0.56 (equal to the highest acceptance of the excited state signals) over the entire `other backgrounds' spectrum to mimic this effect. Under these assumptions, the S/N before applying KamNet is 0.0691, and the S/N after applying KamNet is 0.1187, which corresponds to a 72\% improvement. 

\section*{Conclusion and Outlook}
\label{sec:conclusion}
Liquid scintillator detectors have been at the heart of many of the great discoveries in neutrino physics and have been a leading technology in the search for \vbb-decay. Their data is effectively a time series of images projected onto a sphere.  In our previous work~\cite{nim_paper}, the power of conventional CNN models had been demonstrated. In this work, we invented a novel deep learning model called KamNet for better performance. Leveraging recent breakthroughs in geometric deep learning and spatiotemporal data modeling, KamNet outperforms the conventional CNN on both rejection efficiency and robustness. With a standard detector configuration similar to the current KamLAND–Zen detector, we find KamNet can reject 74.0\% of the \C~background with 90\% acceptance of the \vbb-decay signal, surpassing 61.5\% rejection from conventional CNN. Furthermore, by applying KamNet to \Xe~ground state and various excited states, we find we can boost the S/N ratio of \Xe~excited state decay search by 72\%. Finally, with precisely tuned MC in KamLAND–Zen 800, we find KamNet can reject 27\% XeLS backgrounds and 59\% film backgrounds without any coincidence tagging or hardware upgrade. We conservatively estimated the \vbb~sensitivity boost from these background rejection to be 10.8\%.

This work has focused on optimizing an algorithm for data from a spherical LS detector, however the data-driven nature of KamNet allows a easy generalization to different detectors and different tasks. Furthermore, the network interpretation study we performed unravels the black-box nature of KamNet to reveal underlying physics. Our future work moves in two directions. We intend to perform a systematic interpretation study to rigorously unveil the origin of KamNet's classification power and perhaps further improve the performance of the algorithm. We then plan to extend the reach of KamNet beyond event classification and background rejection. These include but not limited to KamNet-GAN for event generation, Self-supervised KamNet to provide Bayesian posterior distributions for spectrum fitting and Regressive KamNet for event reconstruction. These studies are benefiting from an abundance of work being done for other applications both inside and outside of particle and nuclear physics and this is just the beginning.

\begin{acknowledgments}
This material is based upon work supported by the National Science Foundation under Grant Numbers 2110720, 2012964, and the U.S. Department of Energy, Office of Science, Office of Nuclear Physics, under Award Number A22-0804-001. This work is done in support of the KamLAND–Zen experiment and we thank our collaborators for their input. The KamLAND-Zen experiment is supported by JSPS KAKENHI Grant Numbers 21000001, 26104002, and 19H05803; the Dutch Research Council (NWO); and under the U.S. Department of Energy (DOE) Grant No. DE-AC02-05CH11231, as well as other DOE and NSF grants to individual institutions. We thank Prof. Julieta Gruszko for the support of Aobo Li's work on KamLAND–Zen. This research was performed, in part, using the Boston University Shared Computing Cluster.
\end{acknowledgments}


\bibliography{kamnet_paper}

\end{document}